\documentstyle[12pt]{article}
\makeatletter \ifcase\@ptsize \font\teneufm=eufm10
\font\seveneufm=eufm7 \font\fiveeufm=eufm5
\font\teneusm=eusm10 \font\seveneusm=eusm7
\font\fiveeusm=eusm5 \or \font\teneufm=eufm10 scaled
\magstephalf \font\seveneufm=eufm7 \font\fiveeufm=eufm5
\font\teneusm=eusm10 scaled \magstephalf
\font\seveneusm=eusm7 \font\fiveeusm=eusm5 \or
\font\teneufm=eufm10 scaled \magstep1 \font\seveneufm=eufm7
\font\fiveeufm=eufm5 \font\teneusm=eusm10 scaled \magstep1
\font\seveneusm=eusm7 \font\fiveeusm=eusm5 \fi

\newfam\eufmfam \newfam\eusmfam \textfont\eufmfam=\teneufm
\scriptfont\eufmfam=\seveneufm
\scriptscriptfont\eufmfam=\fiveeufm
\textfont\eusmfam=\teneusm \scriptfont\eusmfam=\seveneusm
\scriptscriptfont\eusmfam=\fiveeusm

\def\frak{\ifmmode\let\next\frak@\else
 \def\next{\errmessage{Use \string\frak\space only in math
 mode}}\fi\next} \def\frak@#1{{\frak@@{#1}}}
 \def\frak@@#1{\fam\eufmfam#1} 
 \def\sh{\ifmmode\let\next\sh@\else
 \def\next{\errmessage{Use \string\sh\space only in math
 mode}}\fi\next} \def\sh@#1{{\sh@@{#1}}}
 \def\sh@@#1{\fam\eusmfam#1}

\ifcase\@ptsize \font\tenmsa=msam10 \font\sevenmsa=msam7
 \font\fivemsa=msam5 \font\tenmsb=msbm10
 \font\sevenmsb=msbm7 \font\fivemsb=msbm5 \or
 \font\tenmsa=msam10 scaled \magstephalf
 \font\sevenmsa=msam7 \font\fivemsa=msam5
 \font\tenmsb=msbm10 scaled \magstephalf
 \font\sevenmsb=msbm7 \font\fivemsb=msbm5 \or
 \font\tenmsa=msam10 scaled \magstep1 \font\sevenmsa=msam7
 \font\fivemsa=msam5 \font\tenmsb=msbm10 scaled \magstep1
 \font\sevenmsb=msbm7 \font\fivemsb=msbm5 \fi

\newfam\msafam \newfam\msbfam \textfont\msafam=\tenmsa
\scriptfont\msafam=\sevenmsa
\scriptscriptfont\msafam=\fivemsa \textfont\msbfam=\tenmsb
\scriptfont\msbfam=\sevenmsb
\scriptscriptfont\msbfam=\fivemsb

\def\Bbb{\ifmmode\let\next\Bbb@\else
 \def\next{\errmessage{Use \string\Bbb\space only in math
 mode}}\fi\next} \def\Bbb@#1{{\Bbb@@{#1}}}
 \def\Bbb@@#1{\fam\msbfam#1} \def\hexnumber@#1{\ifnum#1<10
 \number#1\else \ifnum#1=10 A\else\ifnum#1=11
 B\else\ifnum#1=12 C\else \ifnum#1=13 D\else\ifnum#1=14
 E\else\ifnum#1=15 F\fi\fi\fi\fi\fi\fi\fi}
 \def\msa@{\hexnumber@\msafam} \def\msb@{\hexnumber@\msbfam}
 \mathchardef\square="0\msa@03

\makeatother
\newcommand{\beq}{\begin{equation}}
\newcommand{\eeq}{\end{equation}}
\newcommand{\ba}{\begin{array}}
\newcommand{\ea}{\end{array}}
\newcommand{\bea}{\begin{eqnarray}}
\newcommand{\eea}{\end{eqnarray}}
\newcommand{\bean}{\begin{eqnarray*}}
\newcommand{\eean}{\end{eqnarray*}}
\newtheorem{theorem}{Theorem}[section]

\newtheorem{remark}[theorem]{Remark}

\newtheorem{proof}{Proof.}

\makeatletter \@addtoreset{equation}{section}

\makeatother
\newcommand{\HH}{{\Bbb H}} \newcommand{\RR}{{\Bbb R}}
\newcommand{\CC}{{\Bbb C}} \newcommand{\PP}{{\Bbb P}}

\newcommand{\DD}{{\Bbb D}}
\def\th{\theta}  
\def\Ga{\Gamma} \def\be{\beta} \def\al{\alpha}

 \def\sig{\sigma} \def\Sig{\Sigma}

\newcommand{\ajm}[3]{Amer. J. Math. {\bf #1} (#2), #3}
\newcommand{\cmp}[3]{Comm. Math. Phys. {\bf #1} (#2), #3}
\newcommand{\pl}[3]{Phys. Lett. {\bf B #1} (#2) #3}
\newcommand{\np}[3]{Nucl. Phys. {\bf B #1} (#2), #3}

\newcommand{\ihes}[3]{Publ. Math. I.H.E.S. {\bf #1} (#2),
#3} 
\newcommand{\ijmp}[3]{Int. Jour. Mod. Phys. A{\bf #1} (#2),
#3} 
\newcommand{\lmp}[3]{Lett. Math. Phys. {\bf #1} (#2), #3}
\newcommand{\plms}[3]{Proc. London Math. Soc. {\bf #1} (#2),
#3} 

\newcommand{\lanl}[1]{LANL preprint, hep-th/#1}
\newcommand{\mpl}[3] {Mod. Phys. Lett. A{\bf #1},(#2),#3}
\newcommand{\rref}[1]{(\ref{#1})} 
\newcommand{\half}{\frac{1}{2}}

  \def\RS {Riemann
surface}\def\RSs {Riemann surfaces}
\newcommand{\VHS}{Variation of Hodge Structure}
\newcommand{\VHSs}{Variations of Hodge Structure}
\def\deriv#1#2{{\strut{\del#1\over\del#2}\displaystyle}}
\def\plusminus{\pm} 
\newcommand{\del}{{\partial}}
\newcommand{\delb}{{\bar\partial}}

\begin{document}
\begin{titlepage}
\begin{flushright}{\AA}rhus Universitet
Matematisk Institut\\ Preprint series 1994 No. 19\\[7pt]
Preprint Montpellier PM/94--38\\[7pt]
hep-th/9411184
\end{flushright}
\vspace{0.5truecm}
\begin{center}
{\huge Toda field theory as a clue\\ \vspace{0.2truecm} to
the geometry of W$_{\rm{n}}$--gravity\footnote{Talk given at
the XI Italian Congress of General Relativity and
Gravitation (SISSA -- Trieste, September 1994)}}
\end{center}
\vspace{0.5truecm}
\begin{center}
{\large\it Ettore Aldrovandi} \footnote{Supported by the
Danish Natural Science Research
Council.}${}^,$\footnote{e--mail: ettore@mi.aau.dk}\\
Matematisk Institut -- \AA rhus Universitet\\ Ny Munkegade
-- DK-8000 \AA rhus C --- DANMARK
\\[0.5truecm]
{\large\it Gregorio Falqui}
\footnote{e--mail: falqui@lpm.univ-montp2.fr}\\
Laboratoire de Physique Math\'ematique \footnote{Unit\'e de
recherche
associ\'ee au CNRS n. 768} ---
Universit\'e Montpellier II \\ Place E. Bataillon, 34095
Montpellier CEDEX 05 --- FRANCE
\end{center}
\vspace{0.1truecm}
\begin{center}
October 1994
\end{center}
\vspace{0.2truecm} \abstract{\noindent We discuss
geometrical aspects of Toda Fields generalizing the links
between Liouville gravity and uniformization of Riemann
surfaces of genus greater than one.  The framework is the
interplay between the hermitian and the holomorphic geometry
of vector bundles on such Riemann surfaces.

Pointing out how Toda fields can be considered as equivalent
to Higgs systems, we show how the theory of Variations of
Hodge Structures enters the game inducing local holomorphic
embeddings of Riemann surfaces into homogeneous spaces.  The
relations of such constructions with previous realizations
of $W_n$--geometries are briefly discussed. }
\end{titlepage}
\setcounter{footnote}{0}

\section{Introduction }
As it was put forward by Polyakov~\cite{Pol81} in his
seminal paper on the geometry of strings, two dimensional
euclidean quantum gravity in the conformal gauge is
described by Liouville theory.  Indeed a fruitful approach
to 2D quantum gravity is to look at it as an effective
theory resulting from minimally coupling conformal matter
$S^0_{mat}(X)$ to metric $g$ on the two--dimensional
universe $\Sigma$.

Integrating out the matter degrees of freedom in the Feynman
path integral, and factoring out the invariance group
$Diff(\Sigma)$ of diffeomorphisms of $\Sigma$ gives rise to
an effective action containing the reparametrization
ghosts and the conformal factor in the metric $g= e^{2\phi} \hat g$.

In such a Faddeev--Popov reduction procedure the
field $\phi$ can be shown to enter the theory through the
Liouville action \beq\label{liouac} S_{Liouv}(\phi;\hat
g)\sim \int_\Sigma d x d t \sqrt{\hat g}(1/2 g^{a b} \del^a
\phi\del^b\phi+\hat R \phi + e^{2\phi})\eeq provided some
reasonable locality assumptions are made~\cite{DDK} when one
takes into account the dependence on the moduli of $\Sigma$.

Following the introduction, originally due to
A. B. Zamolodchikov~\cite{Za85}, of higher spin extensions of the
Virasoro algebra in conformal field theory,
a generalization of ordinary 2D--gravity,
called $W_n$--gravity has recently received a great deal of
attention, also in the attempt of breaking the $c=1$ barrier
characteristic of Liouville theory.

This was both a natural extension of Polyakov's results
about the interpretation of 2D--gravity as a {\em
constrained} $SL(2,\RR)$ WZNW--theory~\cite{Pol90}, and of
the results about matrix models of discretized gravity,
showing their equivalence with the n--KdV equations, whose
symmetry algebra is just the (classical limit of the)
$W_n$--algebra.

Some interesting results in $W_n$--gravity both in the
Feynman path integral approach and in the BRST formalism
have already been
obtained~\cite{DeBGoe,Zu92,Jaya94,GeLeMa91, Hull94}.  In
this paper we want to 
investigate on the geometrical properties of Toda systems
and discuss what are the ``chiral''
embeddings of $\Sigma$ one can associate to a Toda
Field. The main tools are the introduction of some auxiliary
vector bundles over $\Sigma$ (i.e. making a detour through
gauge theory) and the study
of the interplay of their
differential and holomorphic geometry.  We will stick mainly
to the $A_n$ Toda case, and consider compact \RSs\ of genus
greater than one, a first and easier step in the study of
the more general case of negatively curved pointed surfaces.

Actually, as we will clarify in section~\rref{liouv}, we
follow a path suggested by Liouville theory, and argue that
as Liouville theory is the classical Poincar\'e -- Koebe --
Klein uniformization of \RSs, the geometrical structure
behind $W_n$--gravity is the ``higher--order
uniformization'' of Hitchin, Simpson and others, i.e.  the
geometry of Higgs bundles over $\Sigma$, structures that
were first introduced by Hitchin~\cite{Hit87} in the
framework of the self--dual Yang Mills equations, whose r\^ ole in
this topics is currently under deep investigation~\cite{DeBGoe,Jaya94}.

The starting point the is fact that the Toda equations can
be given the form of zero--curvature equations for a
suitable connection (the {\em Toda connection}), and that
there is a gauge
in which they are equivalent to Hitchin's equation 
for the corresponding Higgs pair.  Making use of the theory
of harmonic Higgs bundles~\cite{Cor88, Simp92} one can
decompose the Toda connection in a {\em metric} part plus a
deformation $\alpha$.  Then the metric gives rise to {\em
harmonic} local maps from $\Sigma$ to a symmetric Riemannian
manifold, which however, is not enough for our purposes
since, in general, the target is {\em not even a complex
manifold}.\\ The analysis can be refined under the light of
the theory of Variations of Hodge Structures, thanks to the
fact that the Toda connection (and so the associated Higgs
system) is quite ``special'' and satisfies the so--called
Griffiths transversality conditions~\cite{Gri84,Simp88}. If
follows that one can associate to any solution of the Toda
equations a holomorphic map of $\Sigma$ into a locally
homogeneous hermitian manifold.

Recalling the intimate relations~\cite{BiGe89} between Toda
theory and $W_n$--algebras, a link between such results and
previous realizations of $W_n$ geometry~\cite{RaSa93,GeSa93}
will be provided by recovering the generalized Pl\"ucker
embeddings associated to the $A_n$ Toda systems.

This paper heavily relies on~\cite{AlFa93}, to which we
refer for missing proofs, a proper mathematical setting and
a more substantial list of references.

\section{Some geometrical aspects of Liouville theory}\label{liouv}
In this section we will recall some facts about the
Liouville equations with the aim of casting it in a form
suitable to be generalized to that $A_n$ Toda case.

The geometry underlying the Liouville equation is the
classical geometry of uniformization. The Liouville
equations obtained from the action~\rref{liouac} are, in
complex coordinates $z=x+ \sqrt{-1}t ,\> \bar
z=x-\sqrt{-1}t$, \beq \del\delb \phi=e^{2 \phi}\eeq They are
consistent if and only if the Liouville mode $\phi$ is
regarded as the conformal factor in a metric $g$ over
$\Sigma$ \beq g_{z\, \bar z}=e^{2 \phi} d z\otimes d \bar
z\eeq and can be expressed as the condition for the
constancy of the curvature scalar of $g$.

The universal \RS\ supporting its standard solution is the
complex upper half plane $\HH$ endowed with the Poincar\'e
metric \beq d s^2=e^{2 \phi} d z\otimes d \bar z=
{{1}\over{({\rm Im} z)^2}} d z\otimes d \bar z\eeq Since
this is invariant under projective transformations
\beq
z\to{{a z+b}\over{c z +d}},\>\left(
\begin{array}{cc}
  a & b\\ c & d\end{array}\right) \in SL(2,\RR )
\eeq
one can start by finding solutions (i.e. defined in an open
simply connected coordinate patch of $\Sigma$) by pull-back
as \beq e^{2 \phi}d z\otimes d \bar z={{\del A\delb
    B}\over{(A-B)^2}}d z\otimes d \bar z\eeq with $\delb
A=\del B=0$.  Then reality $\overline{e^{2 \phi}}=e^{2
  \phi}$ is enforced by requiring that $B(\bar z)$ be at
most a real projective transformation of $\overline{A(z)}$,
thereby inducing a real structure. We shall see real
structures play a major role in the sequel.\\ To proceed
further, let us define \bea T(z)=e^{\phi}\del^2e^{- \phi}\\
\bar T(\bar z)=e^{\phi}\bar\del^2e^{- \phi}\eea Then, with
the usual notation of $\{f,z\}$ for the Schwarzian
derivative of $f$ with respect to $z$, $T=-1/2\{A,z\}$ and
$\bar T=-1/2\{B,\bar z\}$, so that by putting
$\xi_1={{1}\over{\sqrt{A}}},\> \xi_2={{A}\over{\sqrt{A}}}$,
it is easy to check that $\xi_{1,2}$ are a basis of the
space of solutions of the equation \beq\label{teq}
-\del^2\xi+T(z)\xi=0\eeq To globalize such local solutions
to a non--trivial \RS\ $\Sigma$ we consider the local system
$\Xi_\alpha=(\xi_{1_\alpha},\xi_{2_\alpha})$, which
solves~\rref{teq} in the open patch $U_\alpha$, glued in
$U_\alpha\cap U_\beta$ with $\Xi_\beta$ by means
of\beq\label{flatcoc} \Xi_\alpha^i= (k_{\alpha
  \beta})^{-1/2} [S_{\alpha \beta}]^i_j \Xi_\beta\eeq where
$k_{\alpha \beta}=\left({{d z_\beta}\over{d
    z_\alpha}}\right)$ is the $\CC^*$--cocycle defining the
canonical bundle and $[S_{\alpha \beta}]$ is a {\em flat}
$SL(2,\CC)$ cocycle.  Associated to such a bundle
$K^{-1/2}\otimes S$ we can consider the jet bundle of
$-1/2$--differentials. This arises in the following way.

Let us differentiate the relation~\rref{flatcoc} with respect
to $z_\be$.  We get \beq ({{d}\over{z_\beta}}\,k_{\alpha
\beta}^{1/2}) \Xi_\alpha(z_\alpha)+ (k_{\alpha
\beta})^{-1/2}{{d \Xi_\alpha}\over{d z_\alpha}}=[S_{\alpha
\beta}] {{d \Xi_\beta}\over{d z_\beta}}\eeq a relation which
shows that the one--cochain $F_\alpha=(\Xi_\alpha^\prime,
\Xi_\alpha)$ intertwines between the flat cocycle $S_{\alpha
\beta}$ and the cocycle \beq\label{jetcoc} J^1_{\alpha
\beta}=\left(\begin{array}{cc} k_{\al\be}^{1/2} &0 \\
\del_\be \log k_{\al\be}
&k_{\al\be}^{-1/2}\end{array}\right)\eeq An extensive
analysis of such a representation for Liouville theory has
been performed in~\cite{AlBo93}.  What is relevant for us is
the following picture.

Let us consider the ($C^\infty$) vector bundle
$E=K^{-1/2}\oplus K^{1/2}$; a result which has long been
known in the literature~\cite{LeSa} is that the Liouville
equations are the zero--curvature conditions for the
connection $\nabla_A$ on $E$ defined by
\beq\label{tocon} d_A^\prime=\del+\left(\begin{array}{cc}
\del\phi &1\\ 0 &-\del\phi\end{array}\right),\qquad
d_A^{\prime\prime}=\delb+\left(\begin{array}{cc} 0 &0\\
e^{2\phi}
&0\end{array}\right)\eeq

For our purpose, it is crucial to analyze further this
structure and point out some facts.  First of all, the
term $e^{2 \phi}$ appearing in the lower left corner of the
connection matrix $A_{\bar z}$ has the right covariance
properties to be interpreted as a metric $g$ {\em on}
$\Sigma$, since it is a section of
$Hom(K^{-1/2},K^{1/2})\otimes \bar K\simeq K\otimes \bar K$,
or in other words is a $(1,1)$--form; also
\beq\omega=\del\phi=1/2\del\log g\eeq is the {\em metric}
connection on $K^{-1/2}$ relative to the fiber metric
$\sqrt{g}$ (and $-\del\phi$ is the fiber metric on
$K^{1/2}$).

Furthermore, the holomorphic fiber bundle supporting the
connection $\nabla_A$ is the bundle\beq\label{jet1/2} 0\to
K^{1/2}\to E_2\to K^{-1/2}\to 0\eeq of 1--jets of
$-1/2$--holomorphic differentials whose representing cocycle
is given by ~\rref{jetcoc}, and under the gauge
transformation $\gamma=\left(\begin{array}{cc} 1 &0\\ \omega
&1\end{array}\right)$ the connection $\nabla_A$ is
transformed into the {\em analytic} one $\nabla_W$ which
reads\beq\label{dscon}
d_W^\prime=\del+\left(\begin{array}{cc} 1 &0\\
-\del\omega+\omega^2 &1\end{array}\right),\quad
d_W^{\prime\prime}=\delb\eeq A couple of remarks are in
order: from the physical point of view, one can notice that
since $\omega=\del\phi$, we have recovered the improved
energy momentum tensor $T=(\del\phi)^2-del^2\phi$
as the only
non--vanishing component of the analytic connection in the
Drinfel'd--Sokolov (or $W$)--form.

{}From the mathematical point of view, there is no
contradiction about the existence of an analytic flat
connection on $E_2$, since it is a flat irreducible
holomorphic bundle (and, moreover, the unique rank 2 bundle
admitting a filtration as in~\rref{jet1/2}), as it can be
seen that its extension class is $1/2$ of the Chern class of
$\Sigma$ which is non vanishing as long as the curve is
hyperbolic.

The observation which brings Higgs systems into the game is
that the Toda connection~(\ref{tocon}) can be interpreted in
the form $ \nabla_A=\nabla_B+\theta+\theta^*$ where \beq
\nabla_B=\del+\left(\begin{array}{cc} \del\phi &0\\ 0
&-\del\phi\end{array}\right)\quad +\delb\eeq is the metric
connection on $V$ with respect to the fiber metric
$h=\left(\begin{array}{cc} \sqrt{g} &0\\ 0
&1/\sqrt{g}\end{array}\right)$ (recall that $g=e^{2 \phi}$,
$\theta=\left(\begin{array}{cc} 0 &1\\ 0 &0\end{array}\right)
dz$ is a (1,0)--form with values in $End(V)$ and $\theta^*$ is
the {\em metric adjoint} of $\theta$).  The zero curvature
(alias Toda) equations $F(A)=0$ are then traded for
the equations
\bea\label{hieq}
F(B) &=& - [\theta,\theta^*]\\
\delb \theta &=& 0
\eea
stating that actually $\theta$ is
holomorphic, and that $(E, \theta)$ constitute a stable Higgs
pair~\cite{Hit87}.  This will be the form of Liouville
theory we are going to discuss in the $n$--component Toda
case.

\section{Higgs bundles, Harmonic bundles, and the Toda
  equations}
\label{higgs}

Let $\Sigma$ be a genus $g$ \RS\ with canonical bundle $K$.
A Higgs bundle over $\Sigma$ is a pair $(E,\theta)$ with E a
holomorphic rank $n$ vector bundle over $\Sigma$ and
$\theta$ a holomorphic $End(E)$--valued $(1,0)$ form.  A
Higgs bundle is stable if the slope (i.e. the ratio of the
first Chern class
with the rank)
of every non--trivial {\em $\theta$ -- invariant} subbundle
$F\subset E$ is less than the slope of $E$ itself.

The generalization of Narasimhan -- Seshadri uniqueness
theorem for stable Higgs pairs states
that~\cite{Hit87,Simp88} if $(E,\theta)$ is stable and
$c_1(E) = 0$ there is a unique unitary connection $\nabla_H$
compatible with the holomorphic structure, such that
\begin{equation}\label{nigel}
F_H +[\theta,\theta^*]=0
\end{equation}

The notion of Higgs system can be related to the one of {\em
harmonic bundle}.  Let $V$ be a complex rank $n$ vector
bundle equipped with a flat connection $\nabla$.  The
introduction of an hermitian fiber metric $H$ on $V$ amounts
to a reduction of the structure group to the unitary
subgroup, and allows for a splitting of the connection as
\begin{equation}
\nabla = \nabla_{\! H} + \al
\label{flat2}
\end{equation}
where $\nabla_H$ is a unitary connection and $\al$ is a
1--form with values in the self--adjoint part of
$\mbox{End}(V)$.  The zero--curvature equations for the
connection $\nabla$ are~\cite{Cor88}

\begin{equation}
\nabla_H^2 + \half [\al , \al ] = 0, \qquad\nabla_{\! H}\al
= 0\label{cor1}
\end{equation}

The pair $(V\, ,\nabla )$ is said~\cite{Cor88,Simp92, Don87}
to be {\em harmonic} if we have in addition
\begin{equation}
{\nabla_{\! H}}^{\! *}\,\al = 0
\label{harmonic}
\end{equation}
where the adjoint is taken with respect to a given metric on
$\Sig$.

Since $\al$ is self--adjoint, we can decompose it as $\al =
\th + \th^*$, thus showing that equations (\ref{cor1},
\ref{harmonic}) are equivalent to Hitchin's self--duality
equation \rref{nigel}, supplemented by $\delb\,\th = 0$,
with the $\delb$--operator coming from the (0,1) part of
$\nabla_H$.

Recall that a metric $H$ can be considered as a multi-valued
mapping $f_H:\Sig\rightarrow GL(n,\CC)/U(n)$ or, in other
words, as a section of a bundle over $\Sig$ whose standard
fiber is the coset $GL(n,\CC)/U(n)$. Since
$\nabla$ is flat,
the section $f_H$ can be regarded as a map from the
universal cover of $\Sig$, $f_H:\widetilde{\Sig}\rightarrow
GL(n,\CC)/U(n)$, equivariant with respect to the action of
$\pi_1(\Sig )$, i.e., we have the
commutative diagram
\[
\setlength{\unitlength}{0.0070in}%
\begin{picture}(260,113)(160,540)
\put(175,545){\vector( 1, 0){163}} \put(430,630){\vector(
0,-1){ 65}} \put(160,630){\vector( 0,-1){ 65}}
\put(175,645){\vector( 1, 0){163}}
\put(150,540){\makebox(0,0)[lb]{\raisebox{0pt}[0pt][0pt]{$\Sigma$}}}
\put(350,540){\makebox(0,0)[lb]{\raisebox{0pt}[0pt][0pt]
{$\Gamma\backslash GL(n,\CC)/U(n)$}}}
\put(250,555){\makebox(0,0)[lb]{\raisebox{0pt}[0pt][0pt]
{$\scriptstyle{f_H}$}}} \put(150,640){\makebox(0,0)[lb]
{\raisebox{0pt}[0pt][0pt]{$\widetilde{\Sigma}$}}}
\put(350,640){\makebox(0,0)[lb]{\raisebox{0pt}[0pt][0pt]
{$GL(n,\CC)/U(n)$}}}
\put(250,655){\makebox(0,0)[lb]{\raisebox{0pt}[0pt][0pt]
{$\scriptstyle{f_H}$}}}
\end{picture}
\]
where $\pi_1(\Sig )$ acts on $\widetilde{\Sigma}$ as the
group of deck transformations and on $GL(n,\CC)$ via the
holonomy representation
$\Gamma$.
In the flat coordinate system for $V$ in which $\nabla\equiv
d$ one has
\begin{equation}
\al = -\half f_H^{-1}\, d f_H
\label{alpha}
\end{equation}
which means that $\al$ can be identified with the
differential of $f_H$. Equation \rref{harmonic} implies that
the map $f_H$ is harmonic~\cite{EeSam64} and explains the
terminology.

The link between the Toda equations and Higgs bundles can be
given as follows.  Let ${\frak g}$ be a simple finite
dimensional Lie algebra.  A Toda field over a Riemann
surface $\Sigma$ is a field $\Phi$ taking values in the
Cartan subalgebra ${\frak k}$ of ${\frak g}$ and the Toda
equations
\begin{equation}\label{toda}
\del_z\del_{\bar z}\Phi=\sum h_i \mbox{e}^{\alpha_i(\Phi) }
\end{equation}
can be obtained as the zero curvature equations~\cite{LeSa}
for the connection
\begin{eqnarray}
A_z &=& \half \del_z\Phi + \exp (\half \mbox{ad}\Phi)\cdot
{\cal E}_+\\ A_{\bar z} &=& -\half \del_{\bar z}\Phi +\exp
(-\half \mbox{ad}\Phi)\cdot {\cal E}_-
\end{eqnarray}
where ${\cal E}_+$ (${\cal E}_-$) denote the sum of the
positive (negative) simple roots.

Let us now consider the gauge transformed connection under
the element $g=\exp(\half\Phi)$ \cite{bab89}. We have \beq
A_z^g = \del_z\Phi + {\cal E}_+\qquad A_{\bar z}^g = \exp (-
\mbox{ad}\Phi)\cdot {\cal E}_-\label{T:gauge-2} \eeq We can
now consider $\exp(\Phi)$ as an hermitian form on the
fibers, split $D_A=d+A$ as
\begin{equation}
D_H+\theta+\tilde\theta
\end{equation}
where $D_H$ is the metric connection associated to
$H=\exp(\Phi)$,
\begin{equation}
\theta={\cal E}_+ \mbox{d}z
\label{theta}
\end{equation}
and $\tilde\theta=H^{-1} {\cal E}_- H d \bar z$.  Namely we
have that
\begin{eqnarray*}
{D_H}^\prime &=& \del+\del\Phi\\ {D_H}^{\prime\prime} &=&
\bar\del
\end{eqnarray*}
Notice that $\tilde\theta$ is the metric adjoint
endomorphism of $\theta$. Hence, together with the obvious
fact that ${D_H}^{\prime\prime}\theta=0$, the
zero--curvature equations in this gauge are
\begin{equation}
{D_H}^2+\left[\theta,\theta^*\right]=0
\end{equation}
thus showing that any solution to the Toda equations gives
rise to a well defined
solution of the Hitchin's equations for a suitable Higgs pair.
In particular, if ${\frak g}^\CC=A_{n-1}$, the underlying
vector bundle $E$ is
\begin{equation}
E = \bigoplus_{r = 0}^{n-1} K^{-\frac{n-1}{2}+r}
\label{E:direct}
\end{equation}
The metric $H$ is given by a diagonal matrix whose entries
$h_r=e^{\varphi_r}\, ,\; r=1,\dots ,n$, are themselves
metrics on the factors $K^{-\frac{n-1}{2}+r-1}$ appearing in
\rref{E:direct}.  This completely fixes the transformation
law of the fields $\varphi_r$, (which are related to the
``true'' Toda fields $\phi_i$ by a standard
overparametrization) and one can check that it coincides
with the well--known conformal transformation properties of
the Toda fields \cite{bab89}.

\section{Variations of Hodge Structures}
The purpose of this section is to bring to the light how
solutions of the Toda equations determine local {\em
  holomorphic} maps from $\Sigma$ to a symmetric hermitian
manifold.  These are to be considered as the generalization
of the uniformizing maps associated to local solutions of
the Liouville equation briefly recalled in
section~\rref{liouv}.  The crucial properties for this
analysis are the natural filtration of the Higgs bundle, as
well as the existence of {\em two} real structures, which
enable one to make full use of the theory of Variations of
Hodge structures along the lines of~\cite{Simp92}.

\subsection{The real structures}
\label{add:real}
Consider the basic Higgs system given by \rref{theta} and
\rref{E:direct} together with the harmonic metric $H$.  The
natural extension of Serre duality defines a symmetric
bilinear map $S:E\otimes E\rightarrow{\cal O}_\Sig$
satisfying
\[
S(\th u,v)=S(u,\th v)
\]
for any two local sections $u,v$ of $E$. This can be used
to ensure that in certain representations the structure
group, and hence the holonomy, is reduced to a real form of
$G^\CC=SL(n,\CC)$.  Actually, as found by
Hitchin~\cite{Hit92} it is the adjoint group of the {\em
  split\/} form $G^r =SL(n,\RR)$. It is perhaps of some
interest to notice that the conjugation
$\tau$ in $sl(n,\CC)$ which selects the real form is
concretely given by
\begin{equation}
\tau (\xi )= S\,\bar\xi S,\qquad\xi\in sl(n,\CC)\label{tau}
\end{equation}
where $S$ is the matrix
\begin{equation}
S=\left(\begin{array}{ccccc}
  & & & &1 \\
  & & &\cdot& \\
  & &\cdot& & \\
  &\cdot& & & \\
 1& & & &  \end{array}\right)\label{S}
\end{equation}

We now show there exists another real structure on the
vector bundle $E$.
Let $A:E\rightarrow E$ be the endomorphism equal to $(-1)^r$
on each factor $K^{-\frac{n-1}{2}+r}$. With it, we construct
an indefinite hermitian form $<\cdot ,\cdot >$ over $E$,
namely
\begin{equation}
<u,v>={(A\, u,v)}_H\, ,\qquad u,v\in E
\label{<>}
\end{equation}
A straightforward calculation proves that the hermitian form
\rref{<>} is flat with respect to the Toda connection
$D=D_{\scriptstyle H}+\th +\th^*$, that is we have
\[
\mbox{d}\, <u,v>\, =\, <D\, u,v> + <u,D\, v>\,\qquad u,v\in
E
\]

This implies, of course, that a reduction of the structure
group from $SL(n,\CC)$ to $SU(p,q)$, where $p=[n/2],\,
q=n-p$, takes place. More precisely, what we actually mean
by ``$SU(p,q)$'' is the group corresponding to the fixed
point set in ${\frak g}^\CC =A_{n-1}$ of the conjugation
$\nu$ given by
\begin{equation}
\nu (\xi )= - I\,\rho (\xi )\, I\, ,\qquad \xi\in{\frak
g}^\CC
\label{nu}
\end{equation}
where in this case $\rho$ is simply minus the hermitian
conjugate and $I$ is the matrix having alternatively
$\plusminus 1$ on the principal diagonal
The conjugations $\tau$ defined in~\rref{tau} and $\nu$
commute, so that, upon carefully choosing the
representation,
(the Lie algebra of) the structure group of
the harmonic bundle corresponding to the Toda equations can
be reduced to
the intersection of the fixed point sets of
$\tau$ and $\nu$.  Let us call $G$ the real structure group
so obtained and $K$ its maximal compact
subgroup.

By the results about harmonic bundles quoted in section
\rref{higgs}, we thus obtain a harmonic map
$f_H:\Sig\longrightarrow\Ga\backslash G/K$ and we can
interpret the Toda field equations as the equations
characterizing the embedding of the \RS\ into a some
homogeneous manifold through a harmonic map $f_H$. We can
actually refine this, that is starting from the map $f_H$ we
can define a holomorphic embedding $F:\Sig\rightarrow\DD$
into a {\em complex\/} manifold $\DD$, fibered over $G/K$.
This requires a more extensive analysis of the structure of
the bundle we associated to the Toda equations.

\subsection{Toda systems and Variations of Hodge structures}
Upon rewriting our rank--$n$ basic bundle \rref{E:direct}
as 
\begin{equation}
E=\bigoplus_{r+s=n-1}\,E^{r,s}\, ,\qquad
E^{r,s}=K^{-\frac{n-1}{2}+r}
\label{hodge:1}
\end{equation}
the Higgs field $\th$ appearing in the Toda connection, eq.
\rref{theta}, has the property
\begin{equation}
\th :E^{r,s}\longrightarrow E^{r-1,s+1}\otimes K
\label{hodge:2}
\end{equation}
and the factors are orthogonal with respect to both the
metric $H$ and the indefinite hermitian form $<\cdot,\cdot
>$.  As a consequence, the complete connection $D=D_{\!
H}+\th + \th^*$ satisfies the following {\em Griffiths
transversality condition\/}
\begin{equation}
D:E^{r,s}\longrightarrow A^{1,0}(E^{r-1,s+1})\oplus
A^{1,0}(E^{r,s})\oplus A^{0,1}(E^{r,s})\oplus
A^{0,1}(E^{r+1,s-1})
\label{hodge:3}
\end{equation}
where by $A^\bullet (E^{r,s})$ we mean $C^\infty$
sections. It is useful for later purposes to rewrite
\rref{hodge:3} in the following form. Consider the
filtration \beq\label{filtrEF} E\equiv F^0\supset
F^1\supset\cdots\supset F^{n-1}\supset F^{n}\equiv \{ 0\}
\eeq where \beq
\label{filter}
F^q=\bigoplus_{r=q}^{n-1}\,
K^{-\frac{n-1}{2}+r}=\;\bigoplus_{r=q}^{n-1}\, E^{\, r,s}
\eeq Then the transversality condition can be restated as
\begin{eqnarray}
& &D^\prime :F^q\longrightarrow A^{1,0}(F^{q-1})\nonumber\\
& &D^{\prime\prime}:F^q\longrightarrow A^{0,1}(F^q)
\label{hodge:4}
\end{eqnarray}

According to Simpson, a harmonic bundle
$E=\oplus_{r+s=w}\,E^{r,s}$ whose factors are orthogonal
with respect to an indefinite hermitian form $<\cdot ,\cdot
>$, satisfying~(\rref{hodge:2})
defines a {\em complex variation of Hodge
structure\/} \cite{Simp88,Simp92,Gri84}

It is outside of the scope of this paper to give a complete
introduction to the theory of variations of Hodge
structures; however its basics are the following.

Let us denote by ${\bf E}$ a complex vector space equipped
with
\begin{itemize}
\item a conjugation $\cdot^\sig :{\bf E}\rightarrow {\bf E}$
\item a bilinear form $Q:{\bf E}\times{\bf E}
\rightarrow\CC$ such that:
\begin{enumerate}
\item $Q(v,u)=(-1)^{w}Q(u,v)$, $u,v\in {\bf E}$,
\item it is ``real'' with respect to the conjugation of
${\bf E}$, namely $\overline{Q(u,v)}=Q(u^\sig ,v^\sig )$,
$u,v\in {\bf E}$.
\end{enumerate}
\end{itemize}

A period domain $\DD$ (or, in words, the set of all weight
$w$ Hodge structures on ${\bf E}$) can be defined to be the
set of all (descending) filtrations $\{ {\bf F}^q\}$ in
${\bf E}$ such that
\begin{equation}
\begin{array}{c}
Q({\bf F}^q, {\bf F}^{w-q+1})=0\\ Q(C u, u^\sig )>0
\end{array}
\label{Riem-Gri}
\end{equation}
where $C$ restricts to $\sqrt{-1}^{r-s}$ on each
of the quotients $E^{r,s}=F^r/F^{r+1}$.

Dropping the second condition in the definition, yields the
{\em compact dual\/} $\check{\DD}$ of $\DD$. It is an
algebraic subvariety (actually a {\em manifold}) of a flag
manifold, and hence of a product of Grassmannians
\cite{Gri84}, in which the period domain $\DD$ lies as an
open subset, and therefore as a complex submanifold.

The Higgs bundle
associated to the Toda equations displays the formal
properties of a \VHS\ of weight $w=n-1$ whose subsequent
quotients $E^{\, r,s}=K^{-\frac{n-1}{2}+r}\cong F^r/F^{r+1}$
are line bundles.
Namely on each fibre we have the conjugation $\cdot^\sigma$
as $u^\sigma= S \overline{u}$ (S is given by~(\ref{S})), and the
bilinear form $Q$ is given by~(\ref{<>}).
Applying Griffiths' theory of \VHSs, and
essentially the transversality condition (~\ref{hodge:3})
one proves~\cite{AlFa93}
\begin{theorem}
The Toda equations determine a holomorphic embedding
\[
F_H:\Sig\longrightarrow \Ga\backslash\DD
\]
where $\Ga$ is the monodromy group, $\DD\cong G/K_0$ a
Griffiths period domain, $G$ is the structure group defined
in \S 5.1 and $K_0\subset K\subset G$ a (compact)
subgroup. The map $F_H$ {\em is} the metric $H$ seen as a
section of a flat bundle over $\Sigma$ with typical fiber
$G/K_0$ and its differential is given by the Higgs field
$\th$.
\label{embeddings}
\end{theorem}

\section{Toda fields and W$_n$--geometry}

So far we have analized the correspondence between Toda
field theory and Higgs bundles in the framework of the
theory of hermitian holomorphic vector bundles on a generic
genus \RS\ $\Sigma$. We now come to the last issue,
i.e. ``closing'' the triangular
correspondence
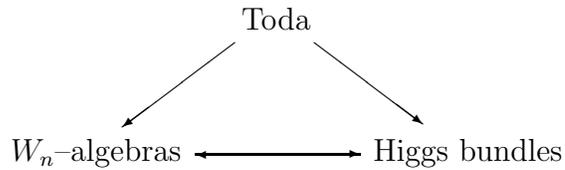
\begin{center}
\setlength{\unitlength}{0.0115in}%
\begin{picture}(176,75)(210,675)
  \put(312,729){\vector(-4,-3){ 51.360}}
  \put(294,678){\vector(-1, 0){ 0}} \put(294,678){\vector(
    1, 0){ 75}} \put(348,729){\vector( 4,-3){ 49.440}}
  \put(375,675){\makebox(0,0)[lb]{\raisebox{0pt}[0pt][0pt]{Higgs
        bundles}}}
  \put(210,675){\makebox(0,0)[lb]{\raisebox{0pt}[0pt][0pt]{$W_n$--algebras}}}
  \put(315,735){\makebox(0,0)[lb]{\raisebox{0pt}[0pt][0pt]{Toda}}}
\end{picture}
\end{center}
{}From the point of view of the theory of connections on
higher genus \RSs\ the relations between Toda Field theory
and $W_n$--algebras is better understood as follows.  The
$(0,1)$ part of any connection $\nabla$ on $\Sigma$ is
integrable by dimensional reasons, thus giving a holomorphic
structure to the complex vector bundle supporting it.  In
this holomorphic frame one has
$\nabla^{\prime\prime}=\delb$.

The characterization of the holomorphic bundle associated to
the basic Higgs bundle \rref{E:direct} and equipped with the
connection (the {\em Toda}--connection)
\begin{equation}\label{T-conn}
  D^\prime = \del + \left(\begin{array}{ccccc}
  \del\varphi_1&1& & & \\ &\del\varphi_2&1& & \\ & &\ddots &
  & \\ & & & &1\\ & & & &\del\varphi_n\end{array}\right) \;
  D^{\prime\prime} =\delb + \left(\begin{array}{ccccc} 0& &
  & & \\ \mbox{e}^{\varphi_1-\varphi_2}&0& & & \\
  &\mbox{e}^{\varphi_2-\varphi_3}&0& & \\ & &\ddots &\ddots
  & \\ & &
  &\mbox{e}^{\varphi_{n-1}-\varphi_n}&0\end{array}\right)
\end{equation}
(here $\sum_{i=1}^{n}\varphi_i=0$), is given~\cite{AlFa93}
by the following
\begin{theorem}
The holomorphic vector bundle $V$ defined by the flat Toda
connection $D=D_{\! H}+\th +\th^*$ {\em is\/} the vector
bundle of $(n-1)$--jets of sections of
$K^{-\frac{n-1}{2}}$. The holomorphic connection $\nabla$,
which is the image of the Toda connection $D$ has the
standard W (or Drinfel'd--Sokolov) form:
\begin{equation}
\nabla^\prime = \del + \left(\begin{array}{cccccc} 0&1& & &
                        & \\ &0&1& & & \\ & & &\vdots & & \\
                        & & & &0&1 \\ w_n & w_{n-1}& &\cdots
                        &w_2 &0
\end{array}\right)\, ,
\quad \nabla^{\prime\prime}=\delb
\label{W:connection}
\end{equation}
with $\delb w_i=0$, $i=2,\dots n$.
\label{Toda-W}
\end{theorem}
The proof of this theorem boils down to show that the vector
bundle $E$, associated in a suitable covering $\{ {\cal
U}_\alpha\}$ of $\Sigma$ by the $SL(n,{\Bbb C})$--cocycle
\beq {\cal E}_{\al\be}=\left(\begin{array}{ccccc}
k_{\al\be}^{{n-1}\over 2}& & & & \\ &
k_{\al\be}^{{{n-1}\over 2}-1} & & & \\ & &\ddots & & \\ & &
& k_{\al\be}^{-{{n-1}\over 2}+1}& \\ & & & &
k_{\al\be}^{-{{n-1}\over 2}}\end{array}\right) \eeq equipped
with the connection $D$ is $C^\infty$--equivalent to the
bundle $V$ of $(n-1)$--jets of sections of $K^{-{{n-1}\over
2}}$, equipped with the connection $\nabla$. We recall that
the transition functions ${\cal V}_{\al\be}$ of $V$ can be
gotten by expanding the relation $\del^l_\al\xi_\al=(
k_{\al\be}^{-1} \del_\be)^l(k_{\al\be}^{{n-1}\over
2}\xi_\be)$,
$\xi_\al$ being a local section
of $K^{ -{{n-1}\over 2}}$, and
$k_{\al\be}=\deriv{z_\al}{z_\be}$.

Rather than dwell at large on the proof, consistent parts of
which were already known or implicit in the literature, we
will examine in some details the $A_2$ case.

The transition functions for the 2--jet bundle of $K^{-1}$
are given by \beq\label{holococ} \left(\begin{array}{c}
\sigma_\al\\ \del_\al\sigma_\al\\
\del_\al^2\sigma_\al\end{array} \right)
=\left(\begin{array}{ccc} k_{\al\be} &0 &0\\ \del_\be \log
k_{\al\be} &1 &0\\ k_{\al\be}^{-2}\del_\be^2 \log
k_{\al\be}&k_{\al\be}^{-1}\del_\be \log k_{\al\be}
&k_{\al\be}^{-1}\end{array} \right)\,
\left(\begin{array}{c}\sigma_\be\\ \del_\be\sigma_\be\\
\del_\be^2\sigma_\be\end{array}\right) \eeq The isomorphism
(in the $C^\infty$--category) between $V=J^2(K^{-1})$ and
$E = K^{-1}\oplus\CC\oplus K$
is accomplished by a {\em smooth} $SL(3,{\Bbb C})$-- valued
$0$--cochain $G_\al$ which we find in the factorized form
$G_\al= G^{(1)}_\al G^{(2)}_\al$, with
\[
G^{(1)}_\al = \left(\begin{array}{ccc} 1 &0 &0\\ h_\al &1
&0\\ f_\al &0 &1\end{array} \right)\qquad G^{(2)}_\al =
\left(\begin{array}{ccc} 1 &0 &0\\ 0 &1 &0\\ 0 &g_\al
&1\end{array} \right)
\]

The Toda connection looks like
\begin{equation}
A_z=\left(\begin{array}{ccc} \del \varphi_1 & 1 &0\\ 0 &\del
\varphi_2 &1 \\ 0 & 0 &\del
\varphi_1\end{array}\right)\qquad A_{\bar
z}=\left(\begin{array}{ccc} 0&0&0\\
\mbox{e}^{\varphi_1-\varphi_2}&0&0\\
0&\mbox{e}^{\varphi_2-\varphi_3}&0
\end{array}\right)\label{TC3}
\end{equation}
Provided we set $h_\al=-(\del\varphi_1)_\al$,
$g_\al=(\del\varphi_3)_\al$, under the transformation $G_\al$
the cocycle~(\ref{holococ}) is sent into its diagonal part
and the Toda connection into its Drinfel'd -- Sokolov
partner \beq (A^{\prime})^{G}=\left(\begin{array}{ccc}
0&1&0\\ 0&0&1\\ w_3&w_2&0\end{array}\right)\, ,\qquad
(A^{\prime\prime})^{G}=0 \eeq
Such a procedure yields
the usual
representation~\cite{BiGe89} for the generators of the $W_3$--algebra:
\beq
w_2=(\del\phi_1)^2+(\del\phi_2)^2-[\del^2\phi_1+\del^2\phi_2+\del\phi_1\del
\phi_2] \eeq and \beq w_3=\del w_2+(\del\phi_1)^2\del\phi_2
-\del\phi_1(\del\phi_2)^2+2\del\phi_2\del^2\phi_2-\del^3\phi_2.
\eeq

Connecting this picture with the results of the previous
sections, we see how datum of a $A_{n-1}$--Toda Field on
$\Sigma$ (and so the datum of a realization of the
$W_n$--algebra) allows to regard the
Riemann Surface $\Sigma$ as a base space for a Variation of Hodge Structure,
and henceforth yields
a holomorphic map from $\Sigma$ into a
quotient of a Griffiths period domain $G/K_0$.

These results are clearly linked to the so called
$W_n$--embeddings as discussed by Gervais, Saveliev and
collaborators, in their works on the geometrical
meaning of the extended local symmetries.
and the eventual
characterization of the $W_n$--moduli spaces.
Although the task is clearly much more difficult than in the
Virasoro case, because of the non--linearity of
$W_n$--algebras, the correct geometrical backgrounds were
pointed out and important steps forwards were made
in~\cite{SoSta91,GeMa92,GeSa93}, where the extrinsic geometry of
``chiral'' embeddings of $\Sigma$ in some projective or
affine space were studied.

In particular, in the paper~\cite{RaSa93} the following
picture is explained. The starting point is a
smooth map from $\Sigma$ to a complex Lie group $G$. Whenever it
satisfies the ``grading condition'', it induces a
holomorphic map $\varphi_P :\Sigma \to G/P$, $P$ being a
parabolic subgroup of $G$. Considering those parabolic
subgroups $P_i, i=1,\ldots ,\mbox{rank}\, G,$ for which
$G/P_i$ is the $i^{th}$ fundamental homogeneous space for
$G$, the associated maps $\varphi_{P_i}$ define maps from
$\Sigma$ to $\PP (V_i)$, the projectivization of the
$i^{th}$ fundamental representation of $G$.  Then it is
shown that the (generalized) Pl\"ucker relations for the
curvature of the pull--back on $\Sigma$ of the Fubini--Study
metrics on $\PP (V_i)$ on $\Sigma$ translate, when expressed
through local K\"ahler potentials, into the Toda Field
equations for a suitably chosen local representative of $
\varphi_{P_i}$.

Our approach can be considered as a
sort of inverse path: we {\em start} from a solution of the
Toda Field equations and we {\em determine} a holomorphic
map from $\Sigma$ to a suitable locally homogeneous space.
It follows that the target space we obtain is only {\em
locally} determined by the rank of the Cartan subalgebra in
which the Toda fields take values, since in the large the
monodromy action of $\pi_1(\Sigma)$ on the
Griffiths period domain must be factored out, thus yielding
a different global target space according to the genus
$g(\Sigma)$.

Nonetheless, Pl\"ucker formulas are of local type, so one
should expect them to arise also in our context. Indeed, one
can argue as follows. Let us overparametrize the Toda fields
$\phi_i$ by
\beq\label{newsurp}\varphi_{-\frac{n-1}{2}+r}=\phi_{r+1}-\phi_r,\qquad
\phi_0=\phi_n=0 \eeq {}which amounts to a renumbering of the
fields $\varphi_i$ entering the Toda connection.  Looking at
the filtration~\rref{filtrEF} one sees that the metric on
$F_q$ is the rank $n-q$ matrix \beq
H_q=H_{|_{F^q}}=\left(\begin{array}{ccc}
e^{\varphi_{-\frac{n-1}{2}+r}}& & \\ &\ddots&\\ & &
e^{\varphi_{\frac{n-1}{2}}}\end{array}\right)\eeq so that
$\mbox{det}
H_q=\exp(\sum_{r=q}^{n-1}\varphi_{-\frac{n-1}{2}+r})$ is a
metric on $\wedge^{max} F^q\equiv\mbox{det} F^q$.  The
defining relations~\rref{newsurp} imply that $\log\mbox{det}
H_q =-\phi_q$, therefore the metric connection on
$\mbox{det} F^q$ is $-\del \phi_q$, and its associated
curvature is $\delb\del \phi_q$.  Writing explicitly the
Toda equations as \beq\label{todaexpl} \delb\del
\phi_q=\exp(2\phi_q-\phi_{q-1}-\phi_{q+1}),\quad
q=1,\ldots,n\eeq we see that the left hand side is the
curvature of $\mbox{det} F^q$, and the right hand side is a
metric on the line bundle $(\mbox{det}
F^q)^{-2}\mbox{det}F^{q-1}\mbox{det}F^{q+1}$.

Recalling that $F^q/F^{q-1}=K^{-(n-1)/2+q}\equiv
\mbox{det}F^q (\mbox{det}F^{q-1})^{-1}$ we get that
$\exp(2\phi_q-\phi_{q-1}-\phi_{q+1})$ is a metric $g_q$ on
$K^{-1}$, i.e. {\em a metric on} $\Sigma$, and then the Toda
equations tell us that $\phi_q$ is the K\"ahler potential
for $g_q$, whose associated 2--form is
$\omega_q=(\sqrt{-1}/2)\,
\del\delb \phi_q$.  The infinitesimal K\"ahler relations
\beq \mbox{Ric}_r=\omega_{r-1}+\omega_{r+1}-2 \omega_r\eeq
follow from standard definitions.

To recover the relations in terms of Pl\"ucker coordinates,
one can take advantage of the naturality of the above
local constructions with respect to the natural embeddings
of the Griffiths' domain $\check{\DD}$ into the product of
Grassmannians~\cite{Gri84}
\[
G(h_1,n)\times G(h_2,n)\cdots \times G(h_{n-1},n)
\]
($h_r$ is the rank of $F^r$ in the Hodge filtration), namely
with respect to the pull backs to $\check{\DD}$ of the
determinant line bundles associated to the tautological
sequences
\[
0\longrightarrow S_{h_r}\longrightarrow
{\CC}^n\longrightarrow Q_{h_r}\longrightarrow 0
\]
over $G(h_r,n)$.

\subsection*{Acknowledgements} We
thank L. Bonora and J--L. Dupont for useful discussions.

\end{document}